# Deep Learning Enabled Design of Terahertz High-Q Metamaterials

Shan Yin, Haotian Zhong, Wei Huang*, Wentao Zhang, Jiaguang Han*

*Guangxi Key Laboratory of Optoelectronic Information Processing,*
*Guilin University of Electronic Technology, Guilin 541004, China*

## ABSTRACT

Metamaterials open up a new way to manipulate electromagnetic waves and realize various functional devices. Metamaterials with high-quality (Q) resonance responses are widely employed in sensing, detection, and other applications. Traditional design of metamaterials involves laborious simulation-optimization and limits the efficiency. The high-Q metamaterials with abrupt spectral change are even harder to reverse design on-demand. In this paper, we propose novel solutions for designing terahertz high-Q metamaterials based on deep learning, including the forward prediction of spectral responses and the inverse design of structural parameters. For the forward prediction, we develop the Electromagnetic Response Transformer (ERT) model to establish the complex mapping relations between the highly sensitive structural parameters and the abrupt spectra, and realize precise prediction of the high-Q resonance in terahertz spectra from given structural parameters. For the inverse design, we introduce the Visual Attention Network (VAN) model with a large model capability to attentively learn the abrupt shifts in spectral resonances, which can efficiently reduce errors and achieve highly accurate inverse design of structural parameters according to the expected high-Q resonance responses. Both models exhibit outstanding performance, and the accuracy is improved one or two orders higher compared to the traditional machine learning methods. Besides, our ERT model can be 4000 times faster than the conventional full wave simulations in computation time. Our work provides new avenues for the deep learning enabled design of terahertz high-Q metamaterials, which holds potential applications in various fields, such as terahertz communication, sensing, imaging, and functional devices.

**KEYWORDS**: High-Q metamaterials, deep learning, abrupt spectral change

# 1. INTRODUCTION

Metamaterials represent a remarkable breakthrough in manipulating of electromagnetic waves. With the unique capability of control polarization[1], wavefront[2] or chirality[3], metamaterials enable widely applications, such as imaging[4], biosensing[4,5], communication[6]. In particular, high-quality (Q) factor metamaterials can significantly enhance light-matter interaction, which have attracted considerable attention[7]. The high-Q devices have great potential applications in optical engineering such as miniaturized lasers[8], nonlinear devices[9], filters[10], and sensors[11].

Traditional methods for manually designing metamaterial-based high-Q functional metamaterial devices involve parameter scanning and repeat simulations during the optimization process. This approach is highly time-consuming, inefficient, and limited, which puts tremendous pressure on time and labor costs. Simultaneously, high-Q metamaterials usually exhibit ultra-narrow resonances in the spectrum. Consequently, in the conventional full wave simulations, the mesh accuracy is extreme high for the computational electromagnetic solver, which further aggravates computational expense. Therefore, a new strategy for efficiently designing high-Q metamaterials is urgently needed.

Machine learning offers an excellent solution for the design of metasurface devices. In recent years, researchers have attempted to combine machine learning with metasurfaces, which mainly applies on inverse design engineering[12,13] and forward prediction of electromagnetic responses[14]. Deep Learning (DL), including Multilayer Perceptron (MLP)[15], Generative Adversarial Network (GAN)[16] or Variational Autoencoders (VAE)[17], has been widely utilized and validated in designing metasurfaces[18–20].

However, deep learning enabled design of high-Q metamaterials is rare. Due to the complex physical processes involved in high-Q metamaterials, even small variations in structural parameters can induce significant changes in spectral resonance, leading to abrupt shifts in resonance frequency and amplitude. Such high sensitivity of variations requires ultra precise mapping relations between the structural parameters and the spectral response, which poses substantial challenges to the learning process both for forward spectral prediction or inverse design engineering. For example, the common method to improve the machine learning model accuracy is to increase the model capacity. But simply enlarging the model capacity will probably introduce a series of issues (such as gradient explosion, gradient vanishing, and network degradation[21]) and result in the invalidation of model. Therefore, it is necessary to develop new model customized for designing high-Q metamaterials.

Here, we propose novel solutions for designing terahertz high-Q metamaterials based on deep learning. To break the limitations of classical machine learning methods, we develop two neural network architectures emphasizing forward prediction of spectral response and inverse design engineering, respectively. For the forward prediction, we utilize the Electromagnetic Response Transformer (ERT) model to establish the complex mapping relations between the highly sensitive variations of structural parameters and the abrupt shifts in spectral resonances, in consequence, we can precisely predict the terahertz spectral response without laborious simulations. For the inverse design, we introduce the Visual Attention Network[22] (VAN) model with a large model capability and the spatial attention focusing on the abrupt shifts in spectral resonances, which can help the neural network better learn the spectral features and thus achieve highly accurate predictions for structural parameters. In addition, we employ transfer learning to improve network training efficiency, and further enhance the accuracy one order higher for the forward spectral predictions and two orders higher for the inverse design compared to the traditional machine learning methods. Besides, our ERT model can be 4000 times faster than the conventional full wave simulations in computation time. Our work provides new avenues for the deep learning enabled design of terahertz high-Q metamaterials, which holds potential applications in various fields, including terahertz communication, sensing, imaging, and more.

## 2. THEORY AND METHODS

Designing high-Q metamaterials generally involves in generating an ultra-narrow resonance in the spectrum, namely, a high Q factor resonance, which can be achieved by quasi-bound state in the continuum[23–26] or Fano resonances through coupling effects[27–30]. The underlying physical mechanisms of high-Q metamaterials are highly complex, which makes the resonance response unpredictable. To manually design high-Q metamaterial devices, it is necessary to analyze each physical process and meticulously optimize each structural parameter. Such work is challenging and time-consuming. Therefore, taking the advantages of deep learning, we aim to achieve high-accuracy prediction of electromagnetic responses and reverse design of high-Q metamaterials without complicated physical process analysis and laborious simulations-optimization.

### 2.1 PHYSICAL MODEL and Training set acquisition

In this study, we select the metamaterials composed of period unit cells of dual split ring resonators (SRRs) as illustrated in Fig.1(a), whose resonance response has been discussed in our previous work[31]. Due to the different coupling and hybridization involving in eigenmodes and lattice mode, the physical process of resonance response is complicated, which leads to the volatile change in spectrum as shown in Fig.1(b). Obviously, the transmission spectra undergo different changes as the angle of incidence increases, and a high-Q resonance will be generated at oblique incidence. In particular, the Q value of the resonance will be influenced significantly by lattice mode (whose frequency is denoted with the red arrows). We then extract the sensitivities of each structural parameter impacting on the transmission spectrum, including period ($P$), width of the SRR ($w$), length of the SRR ($l$), incident angle ($\theta$) and displacement ($d$) as shown in Fig. 1(c). Consequently, the key sensitive parameter can be determined as the incident angle. This result will be an important basis for the subsequent design of neural network architecture.

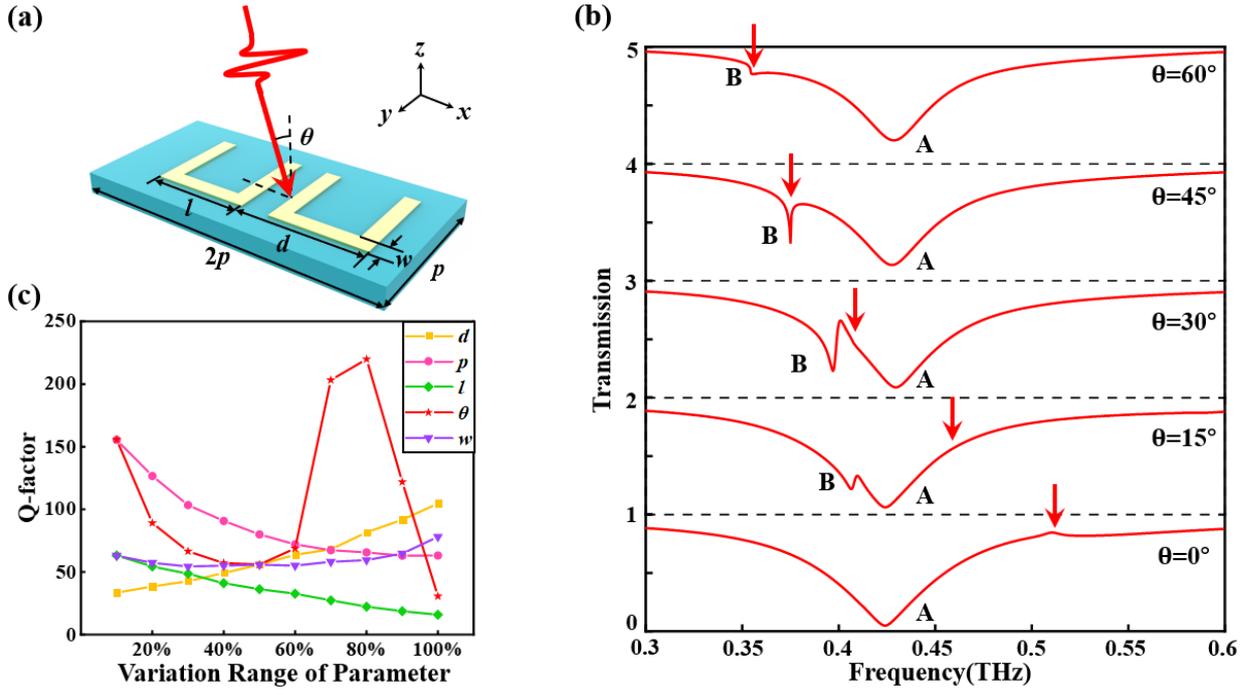

**Fig. 1:** a) Schematic diagram of unit cell of the high-Q metamaterials. (b) Impacts of parameter variation on the Q-factor of peak B. c) Spectral changes in the resonance response as the angle of incidence increases. The red arrows denote the frequency points of the lattice mode.

For the training set, we calculate 9000 transmission spectra using full wave simulations with parameter sweep. The data ranges for each parameter are as follows: $P \in [140,150]\mu m$, $w \in [5,25]\mu m$, $l \in [80,100]\mu m$, $\theta \in [11,60]\mu m$, $d \in [110,130]\mu m$, note that we work with $d$ taking values in the neighborhood of $5/6P$.

## 2.2 DEEP LEARNING MODEL

Due to the abrupt changes in frequency and amplitude in the resonant response of high-Q metamaterials, conventional DL methods have proven ineffective. For example, MLP, a classic machine learning method, has been widely applied in forward prediction and inverse design of metamaterials. However, for metamaterials with high-Q resonance, it becomes necessary to increase the model capacity of neural networks for feature learning. Yet, enlarging the model capacity of MLP introduces a series of issues such as gradient explosion, gradient vanishing, and network degradation[21]. Moreover, other machine learning methods, such as the application of GAN to generate geometric features of high-Q metamaterials, are also faced with challenges. Their accuracy is dependent on numerous intermediate steps, and errors propagated through the generator and discriminator neural network systems can significantly impact on the overall network performance. Additionally, training GAN and their advanced variants presents substantial difficulties, particularly in convergence, which can produce suboptimal results generally in the inverse design of metamaterial. In summary, conventional machine learning methods are no longer sufficient to meet the demands of designing high-Q metamaterials. The intricate physical resonant processes of high-Q metamaterials necessitate more robust deep learning methods.

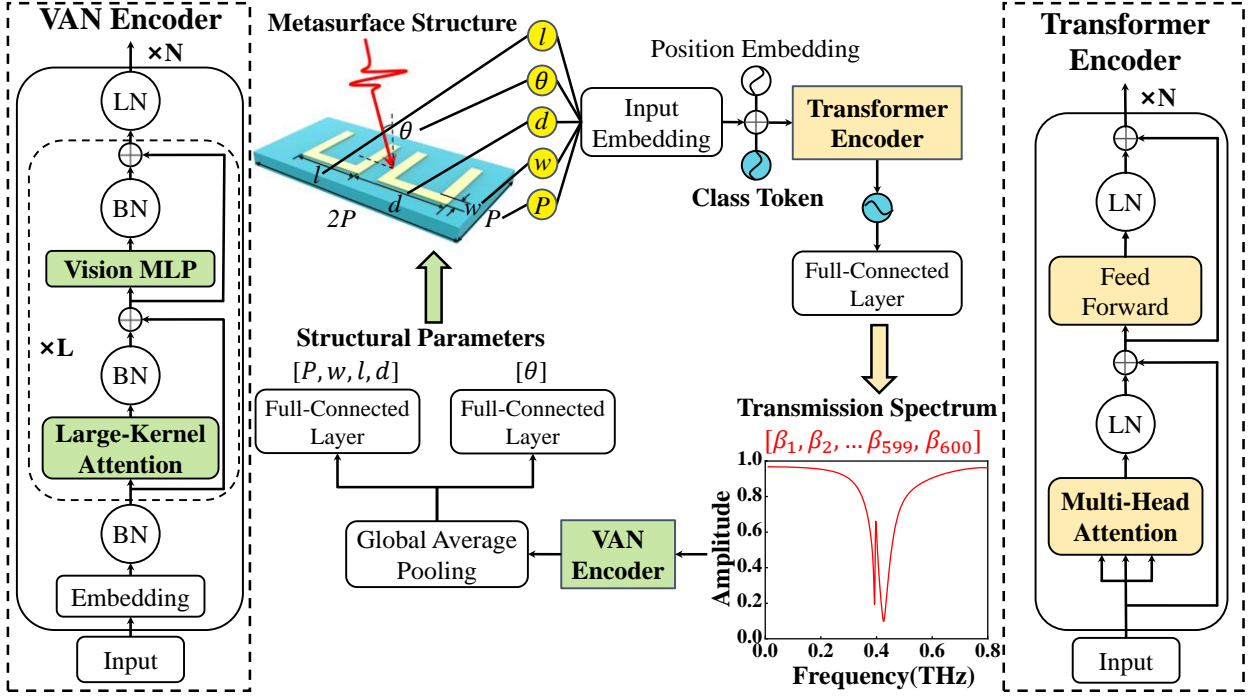

**Fig. 2:** Schematic Diagram of the Workflow for Forward Prediction and Inverse Design Networks

To address these challenges, we propose two neural networks emphasizing forward prediction of transmission spectrum and inverse design of structural parameters as shown in Fig. 2. We utilize the structural parameters of high-Q metamaterials along with their corresponding transmission spectra as a dataset. In our work, five structural parameters are employed to describe the high-Q metamaterials. In the forward prediction task, the structural parameters serve as inputs, while the transmission spectra act as the label, which enables supervised learning. Conversely, in the inverse design task, the inputs and outputs are reversed.

The forward prediction network employs the ERT architecture, primarily consisting of fully connected layers. On the other hand, the inverse design network adopts the VAN architecture, with convolutional layers as the primary components, in particular, we take additional data corrections for the key sensitive parameters and predict them individually, which can efficiently reduce errors and achieve highly accurate predictions for structural parameters. The two networks possess distinct features tailored to their respective tasks. In comparison with classical MLP methods, both network structures exhibit significant model capacity. The ERT model can establish the complex mapping relations between the highly sensitive variations of structural parameters and the abrupt shifts in spectral resonances and the VAN model can learn the abrupt shifts in spectral resonance. As a result, our method significantly enhances both accuracy and efficiency.

## 2.3 FORWARD PREDICTION OF ELECTROMAGNETIC RESPONSE OF HIGH-Q METAMATERIALS

Both the ERT method and classical MLP methods predominantly employ architectures based on fully connected networks. However, the ERT method employs deep learning techniques to perform convolutional high-dimensional feature mapping on the original structural parameter data. It stacks more neural network layers by residual modules, which can effectively mitigate issues such as network degradation, gradient explosion, and gradient vanishing typically associated with stacking neural network layers in MLP methods. Moreover, the ERT method incorporates multiple measures, including multi-head attention mechanisms, positional encoding, and global information embedding, to efficiently learn data features, which allows the ERT method to learn effective feature information about the data distribution within subspaces.

The multi-head attention mechanism[32] in the architecture can be expressed by the following formula:

$$\text{Attention}(Q, K, V) = \text{Softmax}\left(\frac{Q \cdot K^T}{\sqrt{d_k}}\right) \cdot V \qquad (2)$$

where Q represents the input information, forming the query matrix, and K and V appear in pairs as the key matrix and value matrix, respectively. Q, K, and V are obtained through a linear transformation of the input vectors. The specific implementation involves initializing a new matrix randomly, and through iterative training, the input vectors are multiplied by this matrix to obtain new Q, K, and V. The dot product of QK yields the coefficient correlation matrix α. Subsequently, α is divided by the even root of the matrix length and normalized by the softmax function, resulting in the attention weight matrix. The attention weight matrix is multiplied by the V matrix to obtain the attention effect. The attention in each head corresponding to the subspace in the multi-head attention mechanism can be expressed using the following equation.

$$\text{head}_i = \text{Attention}(QW_i^Q, KW_i^K, VW_i^V) \qquad (3)$$

$$\text{MultiHead}(Q, K, V) = \text{Concat}(\text{head}_1, \ldots, \text{head}_n)W^O \qquad (4)$$

The linear transformation in the multi-head attention mechanism is performed H times, where H is the number of heads in the designed architecture. This allows for comprehensive learning of information in its respective subspace.

In comparison to classical prediction methods, the ERT method boasts a larger model capacity and superior prediction accuracy. Hence, we choose the ERT method for the forward prediction of electromagnetic responses in high-Q metamaterials.

### 2.3.1 THE ASSESSMENT OF THE ERT MODEL

We employ the Mean Squared Error (MSE) loss function as the error metric for model evaluation to assess the performance of forward prediction. We compare the performance of the forward ERT model, classical MLP method, and reverse VAN model in their respective forward predictions on the test set, as depicted in Figure 3.

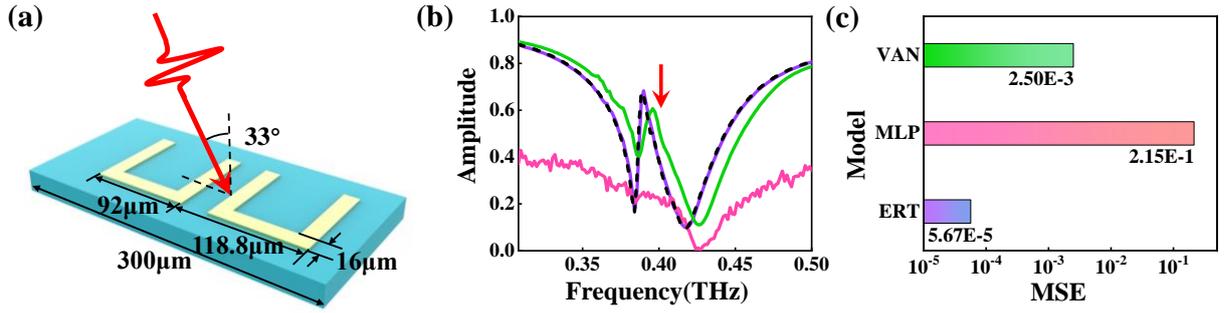

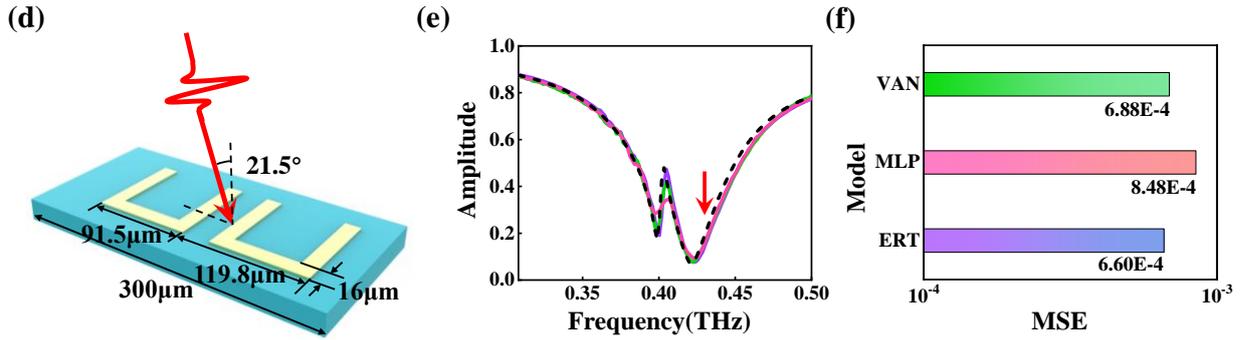

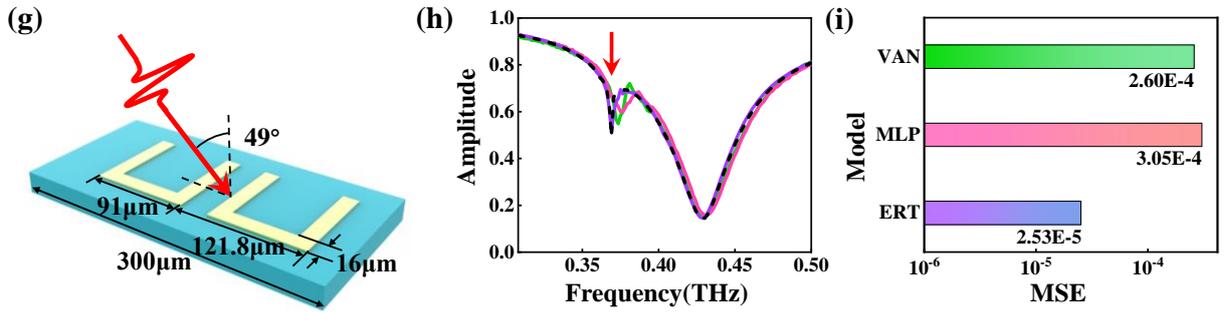

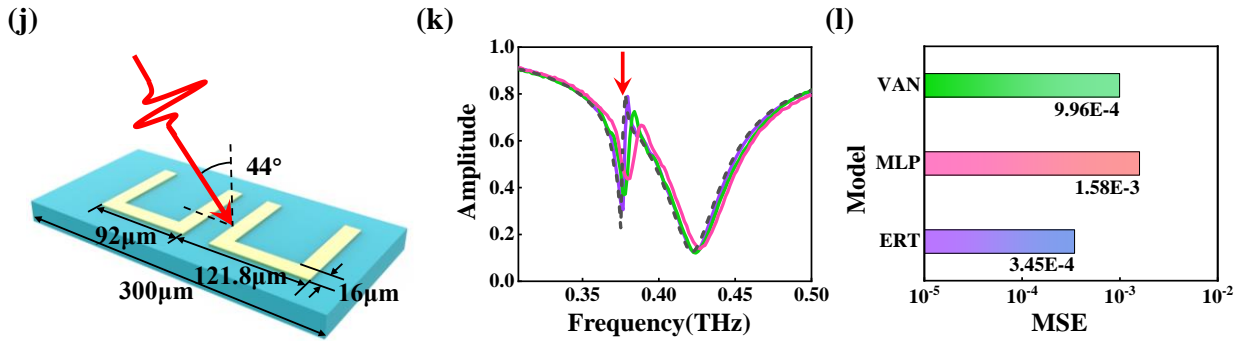

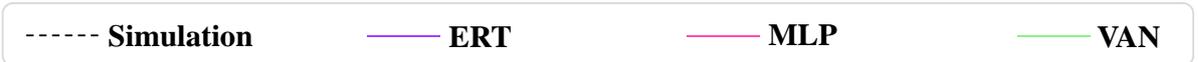

**Fig. 3:** The schematic diagrams of high-Q metamaterial structures under four different scenarios for the ERT model, classical MLP method, and VAN model are illustrated, along with comparisons between predicted and simulated transmission spectra, and MSE between them. A) Randomly select a sample from the test set. B) Interpolating within the parameter range of the test set to ensure the parameters are not outside the entire dataset.

C) Randomly select a sample with a Q value greater than 170 from the test set. D) Extracting the sample with the maximum MSE for the ERT model in the test set.

The figure illustrates schematic diagrams of high-Q metamaterial structures under four different sampling scenarios, along with a comparison of predicted values and true labels for each model, as well as the MSE comparisons. The green, pink, and purple lines represent the predicted spectra of the VAN, MLP, and ERT models, respectively, while the black dashed line represents the labeled(simulated) spectrum for the given structural parameters. The red arrow indicates the frequency of the lattice mode. It can be observed that, in all sampled scenarios, the ERT model consistently exhibits closer proximity to the labeled spectrum compared to other models. The MSE for the ERT model is also consistently lower than that of other models, indicating superior feature learning capabilities in electromagnetic response prediction tasks. The ERT model indeed captures the underlying physical processes of the dataset.

As illustrated in Figure A, the ERT model demonstrates remarkable predictive accuracy for Fano resonances within a parameter range it has not been explicitly trained on. Even in scenarios where the test parameter set involves interpolation within the learned range, as depicted in Figure B, the model maintains accuracy, indicating a continuous rather than discrete learning process. This suggests comprehensive learning of the amplitude abrupt and frequency changes inherent in such physical processes during the training phase. In the instances where the test set exhibits the maximum MSE, the ERT model continues to exhibit high predictive accuracy for both Fano line shapes and intrinsic peaks, as depicted in Figure D. This underscores the model's thorough learning of the features associated with Fano resonances and intrinsic peaks during the training set. Notably, a majority of hybrid modes in the dataset emerge after a 45° incidence angle, before which they manifest as high-Q Fano resonances in the first stage. The ERT method, even with a limited number of hybrid mode samples, excels in accurately predicting the frequency variations and abrupt amplitude changes with minimal error. This proficiency is particularly surprising. It's important to mention that in dataset partitioning, we deliberately selected a portion of the parameter space for the test set that extends beyond the periodicity p range covered in the training set. This choice significantly impacts lattice mode frequencies, which can greatly affect the hybridization mode. This emphasizes the neural network's robust learning of complex physical processes when a decrease in test set loss is observed, validating the generalization ability of the model. Through a comparative analysis of the predictive performance of different models on the test set, the superiority of the ERT model becomes evident. It not only learns Fano resonance responses but also captures the lattice mode induced disruption of Fano peaks and the suppression process of hybrid peaks. In contrast, classical methods fail even in learning the Fano resonance response.

After completing the model performance comparison on highly representative test set samples, we conducted an overall evaluation of the model across the entire test set, as illustrated in Figure 4.

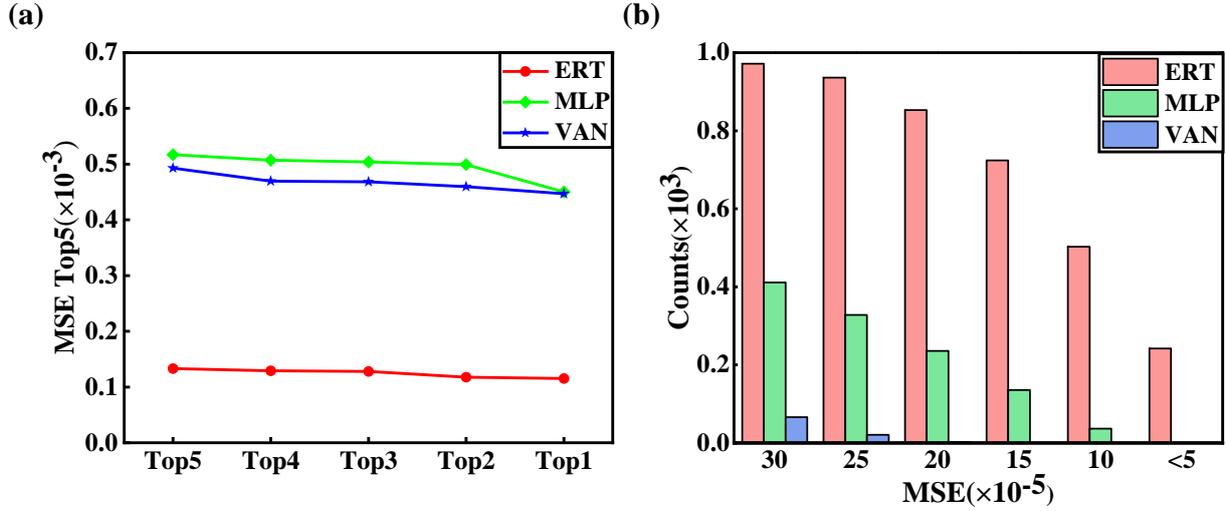

**Fig. 4:** a) Comparison of the TOP5 best average MSEs across the entire test set for ERT, MLP, and VAN models. b) Distribution of prediction errors for the entire test set at optimal accuracy levels for ERT, MLP, and VAN models.

In Figure (a), you can observe the green, blue, and red lines representing the top 5 best MSEs for the MLP model, VAN model, and ERT model, respectively. The results demonstrate that the best testing set MSE for MLP is approximately similar to that of the VAN model. During the VAN model training, feature dimensionality expansion led to a "curse of dimensionality[33]," without feature expansion, VAN loses the advantages of its convolutional methods such as deep convolution and null convolution. These shortcomings contribute to the VAN network achieving MSE values in forward prediction tasks that are even close to those of the classical MLP method, at around $5\times10^{-4}$. Both the classical MLP model and the VAN network model exhibit underfitting tendencies, while the ERT model achieves a remarkable best MSE of $9.4\times10^{-5}$. These outcomes indicate that within the context of electromagnetic response in high-Q metamaterials, classical MLP and VAN models perform poorly in the forward prediction of electromagnetic response. In contrast, the ERT method exhibits exceptional performance, significantly reducing the MSE of the testing set. Figure (b) shows the distribution of MSE samples in the test set under the optimal MSE conditions in Figure (a). The ERT method supports a greater number of low MSE testing instances compared to the MLP method. Among 1000 testing data sets, the ERT model still has 242 samples with an MSE lower than $5\times10^{-5}$, while neither the MLP nor the VAN method has any samples below this MSE threshold.

These findings collectively affirm the capability of the ERT model in learning complex physical processes, enabling accurate predictions of the electromagnetic response of high-Q metamaterials. In contrast, classical MLP methods and the reverse VAN network exhibit underfitting issues, falling short in predicting even Fano resonances. Moreover, they fail to capture the lattice mode induced disruption of Fano peaks and the subsequent suppression of hybrid peaks, rendering them ineffective in providing accurate electromagnetic response predictions for high-Q metamaterials. Consequently, these methods do not meet our design requirements. In contrast, the ERT method excels in feature learning, demonstrating outstanding performance in electromagnetic response prediction tasks. It consistently achieves accurate predictions for high-Q metamaterials, even those with Q values exceeding 150.

## 2.4 INVERSE DESIGN ENGINEERING OF HIGH-Q METAMATERIALS

Compared to traditional inverse design methods of machine learning in inverse design engineering for high-Q metamaterials, VAN stands out when handling data with significant variations, thanks to its notably larger model capacity. This capacity not only allows VAN to map such data into a high-dimensional feature space along channels but also facilitates the extraction and augmentation of features. Beyond this, compared to other convolutional neural networks with large model capacities, VAN's deep convolutional techniques offer superior learning capabilities, enabling more efficient feature mapping and it possesses the advantages of small parameters and computational requirements due to its decomposes large-kernel convolutions into three steps, efficiently capturing long-distance dependency relationships at a low computational cost. This approach proves particularly adept at addressing the challenges associated with the inverse design of high-Q metamaterials.

### 2.4.1 THE ASSESSMENT OF THE VAN MODEL

In this study, considering both model parameter quantity and computational complexity, we selected various convolutional neural networks, the ERT model, the classical MLP model, and the VAN model for the evaluation of inverse design performance. The models were compared based on MSE on the test dataset, and all models eventually converged at four levels of MSE. Therefore, one representative model from each of the four MSE levels was chosen for a detailed evaluation of its performance on the test dataset.

We opted for the MLP model as the classic inverse design method for metamaterial, the classical Resnet model[21] from Convolutional Neural Networks (CNN), and Dual Path Networks[34] (DPN) as experimental subjects to compare with the performance of the VAN model. Initially, we computed the predictions of each model on the test set. Subsequently, the predicted values underwent computer simulation to obtain simulated spectra for each model, which were then compared with the original input transmission spectra. The results are depicted in Figure 5.

## A. Random Sampling of The Test Sets

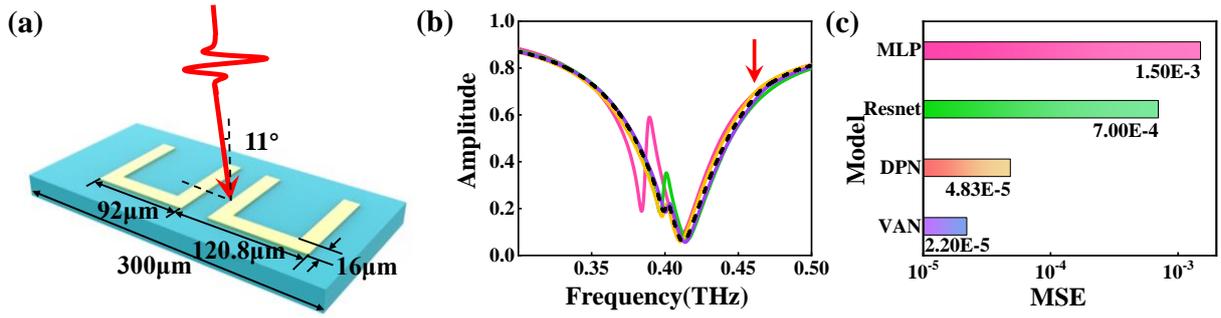

## B. Test Sets Interpolation Sampling

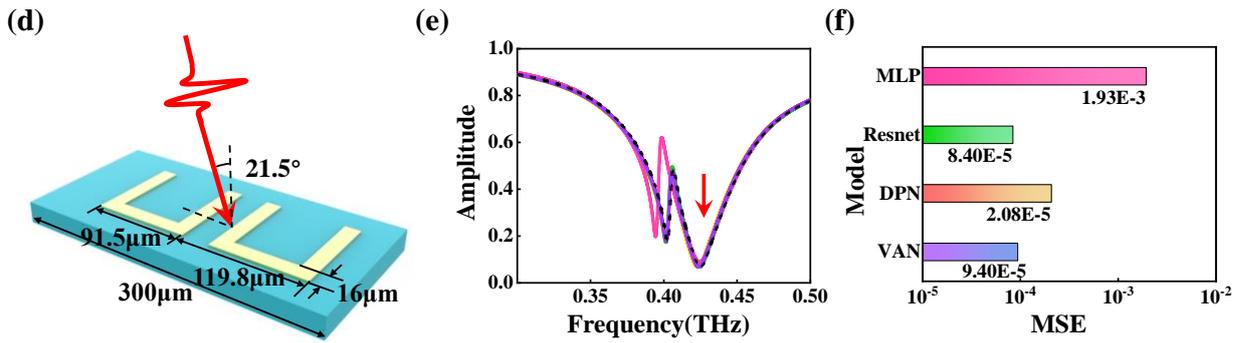

## C. High Q Sample of Test Sets

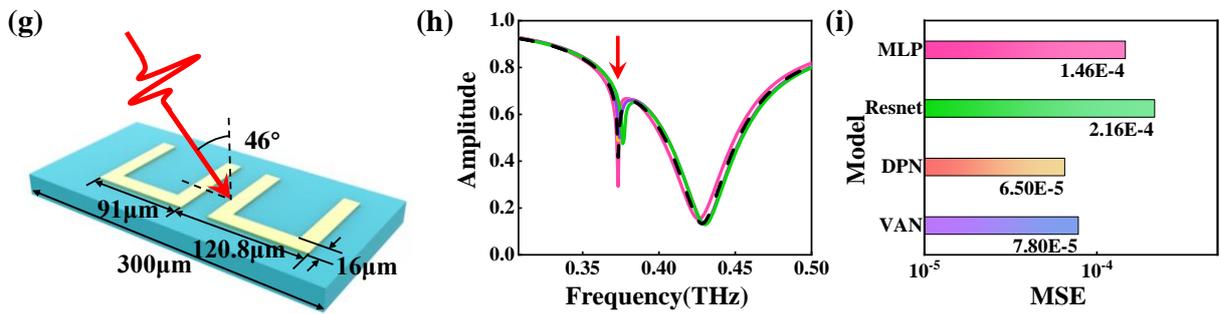

## D. Maximum MSE Sample of The Test Sets

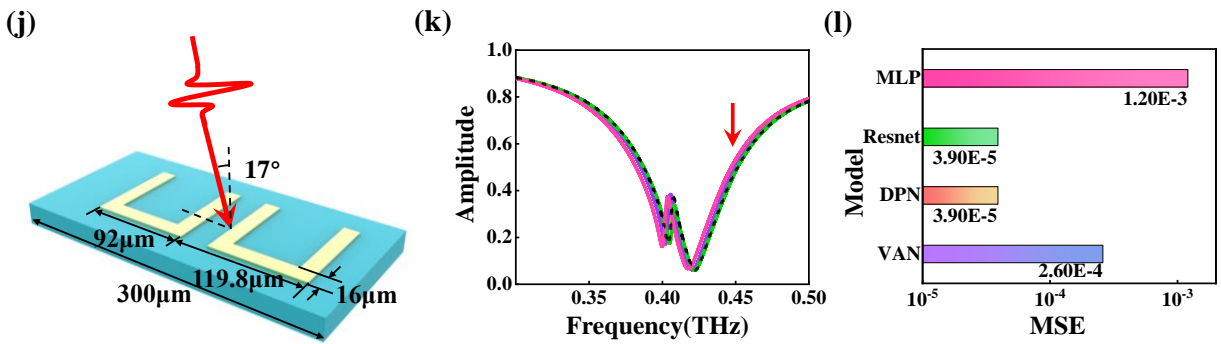

**Fig. 5:** The schematic diagrams of high-Q metamaterial structures under four different scenarios for the four representative models are illustrated, along with comparisons between simulation transmission spectra after predicted and original transmission spectra, and MSE between them. A) Randomly select a sample from the test set. B) Interpolating within the parameter range of the test set to ensure the parameters are not outside the entire dataset. C) Randomly select a sample with a Q value greater than 170 from the test set. D) Extracting the sample with the maximum MSE for the VAN model in the test set.

The specific performance of parameter predictions for four samples is detailed in Table 1.

Table 1. Comparison of model predictions for four scenarios

| Condition | Parameters | Label | VAN | DPN | Resnet | MLP |
|---|---|---|---|---|---|---|
| Random Sampling of The Test Setst | $\theta$ (°) | 11.00 | 10.98(-0.02) | 11.91(+0.91) | 19.55(+8.55) | 30.27(+19.27) |
|  | $p$ (μm) | 300.00 | 298.94(-1.06) | 299.00(-1.00) | 295.79(-4.21) | 300.24(+0.24) |
|  | $l$ (μm) | 92.00 | 92.03(+0.03) | 91.98(-0.02) | 91.82(-0.18) | 92.74(+0.74) |
|  | $w$ (μm) | 15.00 | 15.05(+0.05) | 15.00(+0.00) | 14.90(-0.10) | 15.18(+0.18) |
|  | $d$ (μm) | 120.83 | 120.42(-0.41) | 119.43(-1.40) | 120.09(-0.74) | 119.85(-0.98) |
| Test Sets Interpolation Sampling | $\theta$ (°) | 21.50 | 21.58(+0.08) | 21.66(+0.16) | 22.92(+1.42) | 32.03(+10.53) |
|  | $p$ (μm) | 300.00 | 299.69(-0.31) | 298.78(-1.22) | 297.77(-2.23) | 293.91(-6.09) |
|  | $l$ (μm) | 91.50 | 91.84(+0.34) | 91.80(+0.30) | 91.68(+0.18) | 92.07(+0.57) |
|  | $w$ (μm) | 16.00 | 15.99(-0.01) | 16.04(+0.04) | 15.99(-0.01) | 15.57(-0.43) |
|  | $d$ (μm) | 119.80 | 120.08(+0.28) | 119.88(+0.08) | 119.14(-0.66) | 120.08(+0.28) |
| High Q Sample of Test Sets | $\theta$ (°) | 46.00 | 46.06(+0.06) | 46.42(+0.42) | 44.46(-1.54) | 50.18(+4.18) |
|  | $p$ (μm) | 300.00 | 299.74(-0.26) | 299.49(-0.51) | 299.42(-0.58) | 293.74(-6.26) |
|  | $l$ (μm) | 91.00 | 91.00(+0.00) | 90.99(-0.01) | 90.91(-0.09) | 90.00(+0.00) |
|  | $w$ (μm) | 16.00 | 16.03(+0.03) | 16.01(+0.01) | 15.95(-0.05) | 14.99(-1.01) |
|  | $d$ (μm) | 120.83 | 120.85(+0.02) | 120.57(-0.26) | 120.63(-0.20) | 119.59(-1.24) |
| Maximum MSE Sample of The Test Sets | $\theta$ (°) | 17.00 | 18.11(+1.11) | 17.00(+0.00) | 15.85(-1.15) | 19.53(+2.53) |
|  | $p$ (μm) | 300.00 | 299.25(-0.75) | 299.50(-0.50) | 298.81(-1.19) | 296.21(-3.79) |
|  | $l$ (μm) | 92.00 | 92.27(+0.27) | 92.03(+0.03) | 91.93(-0.07) | 91.99(-0.01) |
|  | $w$ (μm) | 16.00 | 15.91(-0.09) | 16.03(+0.03) | 15.96(-0.04) | 15.04(-0.96) |
|  | $d$ (μm) | 119.83 | 120.14(+0.31) | 120.14(+0.31) | 119.60(-0.23) | 119.27(-0.56) |
| Average Error | \ | \ | 0.2745 | 0.3605 | 1.1710 | 2.9925 |

Similar to the model evaluation work in forward prediction, we continue to compare the performance of the models under four specific sampling scenarios. In the figure, the solid purple, yellow, pink, and green lines represent the spectra predicted by the VAN, DPN, MLP, and Resnet models, respectively, after simulation of the predicted parameters. The black dashed line represents the simulated original spectra, and the red arrows indicate the frequency of the lattice mode. It can be observed that the CNNs perform exceptionally well, with spectra closely approximating the original simulated spectra. In contrast, the classical inverse design method MLP falls slightly behind; for the parameters predicted under the four specific sampling scenarios, there is a considerable discrepancy from the label parameters, and its spectra still exhibit certain differences from the original simulated spectra.

In the inverse design task, the neural network inputs the original simulated spectra and outputs the corresponding parameter sets. This requires the neural network to possess strong learning capabilities to handle the amplitude abrupt and frequency variations in high-Q metamaterials within the dataset. The results indicate that the CNNs have fully learned the characteristics of the strongly mutated dataset, showing outstanding performance. Among the CNNs, although the predictive parameter of Resnet has a small MSE with the parameter labels, in some cases, there is a slight deviation in the shape of the Fano peak in its simulated spectra, as observed in the random sample in Figure A. In contrast, the DPN and VAN models perform excellently in various scenarios. In the random sample, for instance, they accurately predict the parameters when the Fano peak first appears. In the interpolated sample in Figure B, all CNN series networks perform well, indicating that their network learning process is continuous rather than discrete. In the high-Q sample with a Q value of 180 in Figure C, Resnet's simulated spectra show significant deviations in the frequency and amplitude of the hybrid peak, while VAN and DPN also exhibit small amplitude deviations. This highlights that for the design of high-Q metamaterial devices, as the Q value increases, more precise predictions of structural parameters are required. Any minor deviation in frequency and amplitude can impact the Q value, as the matching frequency between lattice modes and Fano resonance is highly dependent on key parameters such as the period and angle. Prediction errors in parameters such as the angle and period have a significant impact on Fano peaks and lattice modes, making the design of high-Q metamaterials highly sensitive to key parameters such as angle and period, requiring high accuracy predictions of parameter values, which we will discuss in more detail later. In the VAN maximum MSE sample in Figure D, even in the case of the largest error for VAN in the test set, it still demonstrates competence in accurate inverse design tasks.

Similar to the forward work, after evaluating these four highly representative samples, we will compare the overall performance of the four models on the test set. We conducted computer simulations on the predicted parameters of the four models across the entire test set, calculating the MSE between the simulated results of the predicted parameters and the original simulated transmission spectra, as shown in Figure 6.

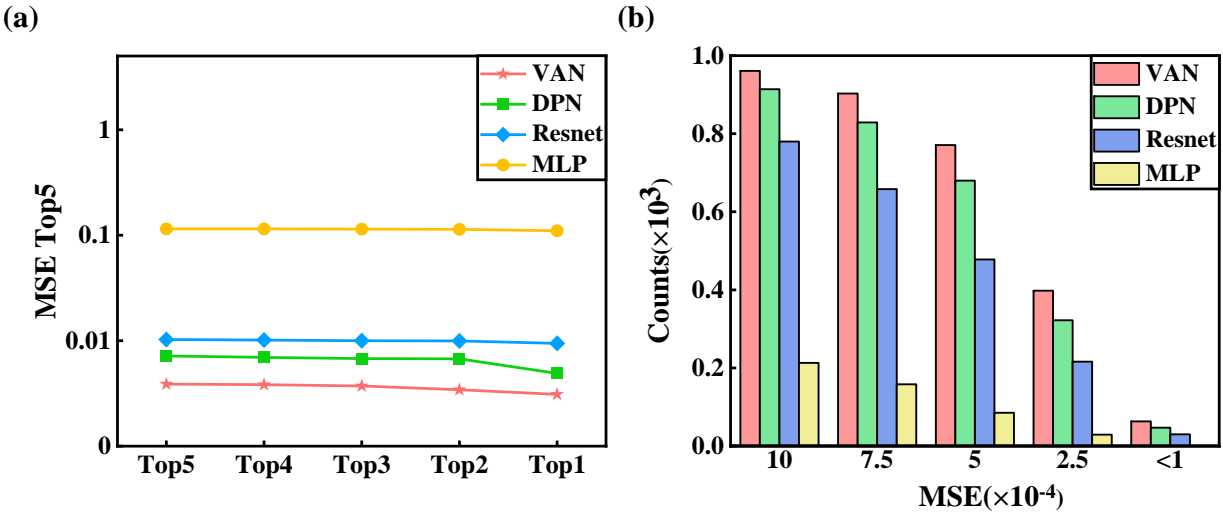

**Fig. 6:** a) Comparison of the TOP5 best average MSEs across the entire test set for VAN, DPN, Resnet, and MLP models. b) The MSE between the transmission spectra simulated by predicted parameters of the four models over the entire test set and the original simulated transmission spectra.

As shown in Figure A, the yellow, blue, green, and red solid lines represent the TOP5 MSE of the test set for the models of MLP, Resnet, DPN, and VAN, respectively, which shows that the VAN model has the best prediction accuracy, with its MSE around 0.003, while DPN comes in second place at around 0.004, and the Resnet is at 0.01, the MLP model has a higher error with its MSE around 0.1. As shown in Figure B, the red, green, blue, and yellow colors correspond to the MSE sample distribution between transmittance spectra after CST simulation for parameter predictions from the VAN model, DPN model, Resnet, and MLP model, respectively, and the original transmittance spectra of the test dataset. The results indicate that the VAN model has the highest number of sample instances in each MSE distribution, which is compatible with the fact that the VAN model has the best MSE among all models in parameter prediction. Following the VAN model, the DPN network, Resnet network, and classic MLP model follow in decreasing order of performance. It is worth noting that MLP and ERT models, characterized by linear layer architectures, are no longer suitable for reverse design engineering, exhibiting larger prediction errors. In high-dimensional spectral data, the convolutional layer demonstrates superior advantages in handling high-dimensional data compared to linear layers, showcasing greater model capacity and better overall performance.

These predictions show that the lower the MSE values predicted by the model for the geometric parameters, the closer the simulated transmission spectra are to the original simulated spectra. This phenomenon stems from the stringent requirement for extreme high accuracy in parameter predictions for high-Q metamaterial devices. The resonant responses of lattice modes and Fano peaks critically depend on two key parameters: periodicity and angle values. While other parameters, such as the distance between double U-shaped rings, also exert some influence, the most significant and sensitive factors are the periodicity and angle values. This is because they collectively determine the matching frequency of lattice modes and Fano peaks, as well as essential features like the Q value and peak shape of Fano resonances, along with the characteristics of hybrid peaks. The more accurate the parameter predictions, the closer the simulated electromagnetic response aligns with the original response, and the closer the Q values approximate.

In the second stage of extreme high-Q hybrid peaks, the larger the Q value of the hybrid peak, the more accurate the frequency prediction of the lattice mode needs to be. The frequency is jointly determined by angle and periodicity parameters. Redshift deviations of lattice modes will lead to the suppression of hybrid peaks in the third stage, resulting in a change in their Q values. Blueshift deviations of lattice modes may alter the frequency matching with Fano peaks, causing significant changes in their Q values. The prediction differences may lead to significant errors in the prediction parameter group of high Q hybrid peaks, and the resonance effect of the predicted parameters after simulation becomes the Fano peak in the first stage. Similarly, during the suppression process of hybrid peaks in the third stage, both redshift and blueshift of lattice modes can significantly impact the resonant effects of hybrid peaks, with a substantial influence on Q values. So predicting angles and periods also requires extreme high accuracy. In the first stage, lattice modes have a minimal impact on spectra. In this case, without considering lattice modes, predicting Fano peaks requires extreme high accuracy in angle values. The key parameter here is the angle value, which affects the resonant effects of Fano peaks, influencing their Q values and resonance frequencies. Considering lattice modes, errors in predicting both periodicity and angle may likely result in frequency matching between lattice modes and Fano peaks, giving rise to hybrid peaks.

In summary, high precision in predicting angle and periodicity values is essential in all three stages. The larger the error, the greater the impact on the design of high-Q metamaterials. The VAN model excels in this reverse design task, demonstrating outstanding performance with high accuracy, minimal computational complexity, and fewer parameters. It effectively accomplishes the reverse design of high-Q metamaterials.

## 3. Conclusion

In this study, we propose a method for predicting the electromagnetic response of high-Q metamaterials based on ERT. This approach automatically predicts the electromagnetic response based on input structural parameters. The reliability of the ERT method is demonstrated through four highly representative samples, and its universality is confirmed through comprehensive testing across the entire dataset. The method achieves precise predictions of electromagnetic responses in various physical processes of high-Q metamaterials, overcoming limitations seen in classical machine learning methods for electromagnetic response prediction.

Additionally, we present a reverse design method for high-Q metamaterials based on the VAN model. The VAN model excels in handling amplitude abrupt in high-Q metamaterials, providing accurate predictions of structural parameters as needed at different physical stages. Through four representative samples under varying scenarios, we illustrate the reliability of the VAN model. Through error comparisons on the test set, we highlight the necessity of high precision in parameter prediction, validating the indispensable role of the VAN model in reverse design for high-Q metamaterials. The VAN model outperforms other CNN models and classical MLP reverse design models, accomplishing reverse design tasks in different physical stages and elucidating key points in reverse design across various physical stages.

These architectures overcome the limited model capacity and further enhance the prediction accuracy in reversely designing the microstructures. We employ transfer learning to improve network training efficiency, achieving precise predictions of electromagnetic response for high-Q metamaterials and on-demand design of high-Q metamaterial structures. These two methods offer novel pathways for the design and electromagnetic response prediction of high-Q metamaterials, holding promise for diverse applications across various fields, including terahertz communication, sensing, imaging, and beyond.

**Funding:** This work was supported by the National Natural Science Foundation of China (62365004, 62235013, 12264010, 62205077), Innovation Project of GUET Graduate Education (2023YCXS231, 2023YCXS229), funding from the Guangxi Oversea 100 Talent Project (W.H.), and funding from the Guangxi distinguished expert project (W.Z, J. H.).

**Notes:** The authors declare no competing financial interest.

# Reference

1. Park J, Kang JH, Kim SJ, Liu X, Brongersma ML. Dynamic Reflection Phase and Polarization Control in Metasurfaces. *Nano Lett*. 2017;17(1):407-413. doi:10.1021/acs.nanolett.6b04378

2. Yin X, Zhu H, Guo H, et al. Hyperbolic Metamaterial Devices for Wavefront Manipulation. *Laser & Photonics Reviews*. 2019;13(1):1800081. doi:10.1002/lpor.201800081

3. Yin S, Chen Y, Quan B, et al. Coupling-enabled chirality in terahertz metasurfaces. *Nanophotonics*. 2023;12(7):1317-1326. doi:10.1515/nanoph-2023-0019

4. Yesilkoy F, Arvelo ER, Jahani Y, et al. Ultrasensitive hyperspectral imaging and biodetection enabled by dielectric metasurfaces. *Nat Photonics*. 2019;13(6):390-396. doi:10.1038/s41566-019-0394-6

5. Wang P, Lou J, Yu Y, et al. An ultra-sensitive metasurface biosensor for instant cancer detection based on terahertz spectra. *Nano Res*. 2023;16(5):7304-7311. doi:10.1007/s12274-023-5386-7

6.  Wu K, Liu JJ, Ding Y jiang, Wang W, Liang B, Cheng JC. Metamaterial-based real-time communication with high information density by multipath twisting of acoustic wave. *Nat Commun*. 2022;13(1):5171. doi:10.1038/s41467-022-32778-z

7.  Wang B, Yu P, Wang W, et al. High-Q Plasmonic Resonances: Fundamentals and Applications. *Advanced Optical Materials*. 2021;9(7):2001520. doi:10.1002/adom.202001520

8.  Tang H, Wang Y, Chen Y, et al. Ultrahigh-Q Lead Halide Perovskite Microlasers. *Nano Lett*. 2023;23(8):3418-3425. doi:10.1021/acs.nanolett.3c00463

9.  Wang C, Fang Z, Yi A, et al. High-Q microresonators on 4H-silicon-carbide-on-insulator platform for nonlinear photonics. *Light Sci Appl*. 2021;10(1):139. doi:10.1038/s41377-021-00584-9

10. Monti A, Alu A, Toscano A, Bilotti F. Design of High-Q Passband Filters Implemented Through Multipolar All-Dielectric Metasurfaces. *IEEE Trans Antennas Propagat*. 2021;69(8):5142-5147. doi:10.1109/TAP.2020.3045795

11. Duan B, Liu S, Liu X, Yu X chong, Wang C, Yang D. High-Q quasi-BIC in photonic crystal nanobeam for ultrahigh sensitivity refractive index sensing. *Results in Physics*. 2023;47:106304. doi:10.1016/j.rinp.2023.106304

12. Chen Y, Lan Z, Su Z, Zhu J. Inverse design of photonic and phononic topological insulators: a review. *Nanophotonics*. 2022;11(19):4347-4362. doi:10.1515/nanoph-2022-0309

13. Li Z, Pestourie R, Lin Z, Johnson SG, Capasso F. Empowering Metasurfaces with Inverse Design: Principles and Applications. *ACS Photonics*. 2022;9(7):2178-2192. doi:10.1021/acsphotonics.1c01850

14. Nadell CC, Huang B, Malof JM, Padilla WJ. Deep learning for accelerated all-dielectric metasurface design. *Opt Express*. 2019;27(20):27523. doi:10.1364/OE.27.027523

15. Tolstikhin I, Houlsby N, Kolesnikov A, et al. MLP-Mixer: An all-MLP Architecture for Vision. Published online June 11, 2021. Accessed December 9, 2023. http://arxiv.org/abs/2105.01601

16. Goodfellow IJ, Pouget-Abadie J, Mirza M, et al. Generative Adversarial Networks. Published online June 10, 2014. Accessed December 9, 2023. http://arxiv.org/abs/1406.2661

17. Kingma DP, Welling M. Auto-Encoding Variational Bayes. Published online December 10, 2022. Accessed December 9, 2023. http://arxiv.org/abs/1312.6114

18. Huang W, Wei Z, Tan B, Yin S, Zhang W. Inverse engineering of electromagnetically induced transparency in terahertz metamaterial via deep learning. *J Phys D: Appl Phys*. 2021;54(13):135102. doi:10.1088/1361-6463/abd4a6

19. Liu Z, Zhu D, Rodrigues SP, Lee KT, Cai W. Generative Model for the Inverse Design of Metasurfaces. *Nano Lett*. 2018;18(10):6570-6576. doi:10.1021/acs.nanolett.8b03171

20. Wang L, Chan YC, Ahmed F, Liu Z, Zhu P, Chen W. Deep generative modeling for mechanistic-based learning and design of metamaterial systems. *Computer Methods in Applied Mechanics and Engineering*. 2020;372:113377. doi:10.1016/j.cma.2020.113377

21. He K, Zhang X, Ren S, Sun J. Deep Residual Learning for Image Recognition. In: *2016 IEEE Conference on Computer Vision and Pattern Recognition (CVPR)*. IEEE; 2016:770-778. doi:10.1109/CVPR.2016.90

22. Guo MH, Lu CZ, Liu ZN, Cheng MM, Hu SM. Visual Attention Network. Published online July 11, 2022. Accessed December 9, 2023. http://arxiv.org/abs/2202.09741

23. Cong L, Singh R. Symmetry-Protected Dual Bound States in the Continuum in Metamaterials. *Advanced Optical Materials*. 2019;7(13):1900383. doi:10.1002/adom.201900383

24. Chen M, Xing D, Su V, Lee Y, Ho Y, Delaunay J. GaN Ultraviolet Laser based on Bound States in the Continuum (BIC). *Advanced Optical Materials*. 2023;11(6):2201906. doi:10.1002/adom.202201906

25. Liu Z, Guo T, Tan Q, et al. Phase Interrogation Sensor Based on All-Dielectric BIC Metasurface. *Nano Lett*. 2023;23(22):10441-10448. doi:10.1021/acs.nanolett.3c03089

26. Zhang F, Chu Q, Wang Q, Zhu S, Liu H. Multiple symmetry protected BIC lines in two dimensional synthetic parameter space. *Nanophotonics*. 2023;12(13):2405-2413. doi:10.1515/nanoph-2022-0781

27. Zhang C, Liu Q, Peng X, Ouyang Z, Shen S. Sensitive THz sensing based on Fano resonance in all-polymeric Bloch surface wave structure. *Nanophotonics*. 2021;10(15):3879-3888. doi:10.1515/nanoph-2021-0339

28. Litvinenko KL, Le NH, Redlich B, et al. The multi-photon induced Fano effect. *Nat Commun*. 2021;12(1):454. doi:10.1038/s41467-020-20534-0

29. Gupta M, Singh R. Toroidal versus Fano Resonances in High $Q$ planar THz Metamaterials. *Advanced Optical Materials*. 2016;4(12):2119-2125. doi:10.1002/adom.201600553

30. Yuan S, Chen L, Wang Z, Wang R, Wu X, Zhang X. Tunable high-quality Fano resonance in coupled terahertz whispering-gallery-mode resonators. *Applied Physics Letters*. 2019;115(20):201102. doi:10.1063/1.5129073

31. Zeng D, Yin S, Guo L, Huang W. High-Q Fano resonance induced by Rayleigh anomaly in terahertz metamaterials. In: *2021 46th International Conference on Infrared, Millimeter and Terahertz Waves (IRMMW-THz)*. IEEE; 2021:1-2. doi:10.1109/IRMMW-THz50926.2021.9567626

32. Vaswani A, Shazeer N, Parmar N, et al. Attention Is All You Need. Published online August 1, 2023. Accessed December 9, 2023. http://arxiv.org/abs/1706.03762

33. Patel NP, Sarraf E, Tsai MH. The Curse of Dimensionality. *Anesthesiology*. 2018;129(3):614-615. doi:10.1097/ALN.0000000000002350

34. Chen Y, Li J, Xiao H, Jin X, Yan S, Feng J. Dual Path Networks. Published online July 31, 2017. Accessed December 9, 2023. http://arxiv.org/abs/1707.01629